\documentclass[preprint,superscript,showkeys,superscriptaddress,amsmath,amssymb]{revtex4-1}

\usepackage{newtxtext,newtxmath}
\usepackage[T1]{fontenc}
\usepackage{graphicx}
\usepackage{bm}
\usepackage{textcomp}
\usepackage{braket}
\usepackage{siunitx}
\usepackage{enumitem}
\usepackage{caption}

\DeclareCaptionLabelSeparator{bar}{ $|$ }
\usepackage[colorlinks=true,
linkcolor=magenta,
citecolor=blue,
pdfauthor={Hyounghan Kwon},
]{hyperref}

\usepackage{times}
\usepackage{dcolumn}
\usepackage{amsmath}
\usepackage[dvipsnames]{xcolor}
\usepackage[labelsep=space]{caption}
\usepackage{setspace}
\usepackage{amssymb}
\usepackage{amsmath}
\usepackage[version=3]{mhchem}
\usepackage{lineno}

\newcommand{\NatureFormat}{%
		\renewcommand{\figurename}{\textbf{Fig.}}
     }
\NatureFormat

\bibpunct{}{}{,}{s}{}{Layout} 
\setcitestyle{super}          
\begin{document}

\title{Scalable Liquid-Crystal Integrated Silicon Nitride Photonic Circuits for Reconfigurable Quantum Interference}


\author{Chunghyun Ahn}
\affiliation{Center for Quantum Technology, Korea Institute of Science and Technology (KIST), Seoul 02792, South Korea}
\affiliation{School of Electrical Engineering, Korea University, Seoul 02841, South Korea}

\author{Yongjin Hwang}
\affiliation{Department of Electrical and Information Engineering, Seoul National University of Science and Technology, Seoul 01811, South Korea}

\author{Sangbaek Lee}
\affiliation{Department of Electrical and Information Engineering, Seoul National University of Science and Technology, Seoul 01811, South Korea}

\author{Jinil Lee}
\affiliation{Center for Quantum Technology, Korea Institute of Science and Technology (KIST), Seoul 02792, South Korea}
\affiliation{Division of Nano \& Information Technology, KIST School, Korea University of Science and Technology, Seoul 02792, South Korea}

\author{Hyunjin Ko}
\affiliation{Department of Electrical and Information Engineering, Seoul National University of Science and Technology, Seoul 01811, South Korea}

\author{Sunghyun Moon}
\affiliation{Center for Quantum Technology, Korea Institute of Science and Technology (KIST), Seoul 02792, South Korea}

\author{Hojoong Jung}
\affiliation{Center for Quantum Technology, Korea Institute of Science and Technology (KIST), Seoul 02792, South Korea}
\affiliation{Department of Physics, Kyung Hee University, Seoul 02447, South Korea}

\author{Hyun-Yong Yu}
\email{Co-Corresponding author: H.J.: yuhykr@korea.ac.kr}
\affiliation{School of Electrical Engineering, Korea University, Seoul 02841, South Korea}
\affiliation{Department of Semiconductor Systems Engineering, Korea University, Seoul, 02841, South Korea}

\author{Se-Um Kim}
\email{Co-Corresponding author: S.-U.K.: seumkim@seoultech.ac.kr}
\affiliation{Department of Electrical and Information Engineering, Seoul National University of Science and Technology, Seoul 01811, South Korea}

\author{Hyounghan Kwon}
\email{Co-Corresponding author: H.K.: hyounghankwon@kist.re.kr}
\affiliation{Center for Quantum Technology, Korea Institute of Science and Technology (KIST), Seoul 02792, South Korea}
\affiliation{Division of Quantum Information, KIST School, Korea University of Science and Technology, Seoul 02792, South Korea}

\maketitle

\textbf{Abstract} 
Integrated quantum photonics requires compact, efficient, and low-power phase modulators. While silicon nitride (SiN) is a promising platform, existing modulators suffer from high power consumption, thermal crosstalk, or high driving voltages. Liquid crystal (LC) offers a compelling alternative because of the large index changes and industrial maturity. However, their suitability for supporting various applications in the photonic quantum system has not been experimentally confirmed. Here, we report the first experimental demonstration that LC-based phase modulators integrated on a SiN platform show highly visible quantum interference. We fabricated a liquid-crystal integrated Mach-Zehnder interferometer (LC-MZI) that achieved CMOS-compatible performance with $V_{\pi} \cdot L < 1\ \text{V}\cdot\text{mm}$. In two-photon interference experiments, the devices exhibited high-visibility quantum interference ($\sim$98.5\%) with voltage-tunable phase control. Furthermore, we validated the scalability of our approach by demonstrating wafer-scale fabrication using stepper lithography. This work establishes LC-integrated SiN photonics as a scalable, reconfigurable, and energy-efficient platform for quantum photonic circuits.

\textbf{Keywords.} Quantum Interference, Integrated Quantum Photonics, Liquid Crystal, Phase modulator, Silicon Nitride, Scalability, Low-Voltage Operation


\section{Introduction}
Quantum photonic technologies hold potential for quantum information processing, quantum communication, and quantum sensing applications~\cite{wang2020integrated,moody20222022,browne2017quantum,slussarenko2019photonic,polino2020photonic,gisin2007quantum}. Quantum photonic systems offer distinct advantages including long coherence, room-temperature operation, and compatibility with existing fiber-based infrastructure. However, realizing practical quantum photonic systems at the chip scale requires the full integration of a wide range of functional components~\cite{wang2020integrated,moody20222022,luo2023recent,psiquantum2025manufacturable,pelucchi2022potential,bao2023very}. In particular, universal quantum photonic processors often rely on networks of tunable interferometers, most commonly Mach-Zehnder interferometers, which demand low-loss and efficient phase modulation~\cite{carolan2015universal,clements2016optimal,wang2018multidimensional,chi2022programmable,bogaerts2020programmable}. In addition, scalable and high-density integration of complex quantum photonic systems can benefit from compact phase modulators that operate at the CMOS level driving voltage. 

Quantum photonic systems inherently require a variety of switching mechanisms tailored to different functionalities~\cite{psiquantum2025manufacturable,aghaee2025scaling}. In practice, no single type of phase modulator can satisfy all these requirements simultaneously, as each involves a trade-off among speed, footprint, power consumption, driving voltage, and scalability~\cite{soref2018tutorial,rahim2021taking,sinatkas2021electro}. For instance, fast electro-optic switches are indispensable for overcoming the probabilistic nature of photon sources and enabling feedforward operations in fault-tolerant and measurement based quantum computing architectures~\cite{prevedel2007high,kaneda2019high,bourassa2021blueprint,bartolucci2023fusion,psiquantum2025manufacturable,aghaee2025scaling}. However, low-loss high-speed modulators on integrated quantum photonic platforms typically rely on Pockels effects, which induce only small refractive index changes. This limitation can lead to larger device footprints, higher driving voltages, and increased electrical packaging complexity, thereby constraining dense integration in large-scale quantum photonic processors~\cite{psiquantum2025manufacturable,aghaee2025scaling,wang2018integrated,abel2019large}. In contrast, many crucial quantum photonic functionalities depend not on ultrafast modulation but on stable, low-power phase control within compact device geometries~\cite{wang2018multidimensional,bao2023very,bartolucci2023fusion,tzitrin2021fault}. Such operations include spectral filtering, pump–seed phase tuning in entangled-photon sources, interferometer stabilization, preparation of photonic entangled states, and the implementation of measurement networks, among others~\cite{psiquantum2025manufacturable,aghaee2025scaling,wang2018multidimensional,chi2022programmable,bao2023very,larsen2025integrated,madsen2022quantum}. In these applications, compactness and low driving voltage potentially take precedence over modulation speed, as they directly dictate integration density, thermal stability, and overall system complexity on the photonic chip. Therefore, practical quantum photonic processors demand phase modulators that combine low-voltage and energy-efficient operation with long-term interferometric stability across large-scale reconfigurable meshes.


On the other hand, silicon nitride (SiN) platforms have emerged as one of the most promising candidates for quantum photonic integrated circuits owing to their CMOS compatibility, low propagation loss, and broad transparency window~\cite{blumenthal2018silicon,harame1987ion,hosseini2014cmos,halir2012ultrabroadband,moss2013new}. Compared to silicon, SiN offers significantly lower optical loss and a wider bandgap, making it particularly attractive for low-noise and high-power-handling quantum photonic applications~\cite{moss2013new,psiquantum2025manufacturable}. However, phase modulators on SiN still face significant challenges. Thermo-optic modulators suffer from high power consumption and thermal crosstalk, which hinder dense integration ~\cite{yong2022power}. Also, strain-optic or electro-optic modulators require complex fabrication and large device footprints, limiting scalability and integration density~\cite{Koen2018,Chrostowski2014,Chen2019,Zhang2021}. Moreover, unless the device length extends to the centimeter scale, these modulators typically require high driving voltages, making direct compatibility with CMOS electronics difficult~\cite{renaud2023sub}.

Liquid-crystal (LC)–based phase modulators have attracted increasing interest as an effective approach to addressing these challenges. These devices leverage intrinsically high birefringence and electro-optic tunability of LC molecules to achieve compact, low-voltage phase modulation with minimal power consumption~\cite{chigrinov2007liquid}. Nematic LCs are predominantly employed in such devices because their simple molecular ordering enables stable alignment even when integrated with complex photonic circuit structures. Unlike thermo-optic approaches, the broad temperature stability range of nematic LCs allows for reliable and efficient room-temperature operation. In addition, the fabrication process of LC-based phase modulators is built upon a mature platform that has been widely deployed across display industries~\cite{yin2022advanced}. Therefore, LC technologies can be integrated into scalable, CMOS-compatible quantum photonic circuits that are well suited for industrial-scale manufacturing. 
Recent efforts have explored LC integration with silicon and SiN platforms, focusing primarily on beam steering, optical switching, and resonator tuning in the telecom and visible spectral regions ~\cite{notaros2022integrated,notaros2023liquid,corsetti2024silicon,dutta2025near}. Despite these seminal advances in LC-integrated programmable photonic devices, LC-based devices operating in the telecom band have not yet been validated for quantum photonic applications. In particular, the integration of LC with SiN platforms has not been demonstrated in this spectral regime, and no studies have explored their suitability for quantum photonics, up to our knowledge. It therefore remains an open question whether LC-based phase modulators can preserve the quantum coherence required for photonic quantum information processing.


In this work, we experimentally validate the compatibility of LC-based phase modulation with quantum photonic operations on the SiN platform. We realized liquid-crystal integrated Mach-Zehnder interferometers (LC-MZIs) that exhibit CMOS-level driving voltages with a compact modulation efficiency of $V_{\pi} \cdot L < 1\ \text{V}\cdot\text{mm}$, where $V_{\pi}$ denotes the voltage required for a $\pi$-phase shift. Two-photon interference measurements confirm that phase tuning through the LC layer maintains indistinguishability and interference visibility, demonstrating its suitability for quantum information processing. Moreover, we establish the scalability of the approach by implementing both electron-beam-defined test devices for quantum experiments and wafer-scale structures patterned via stepper lithography, highlighting its potential for large-scale, reconfigurable quantum photonic circuits.

\section{Results}
\subsection{On-Chip Two-Photon Interference with a LC-MZI}

\begin{figure}
\nolinenumbers
\centering
\includegraphics[width=1\columnwidth]{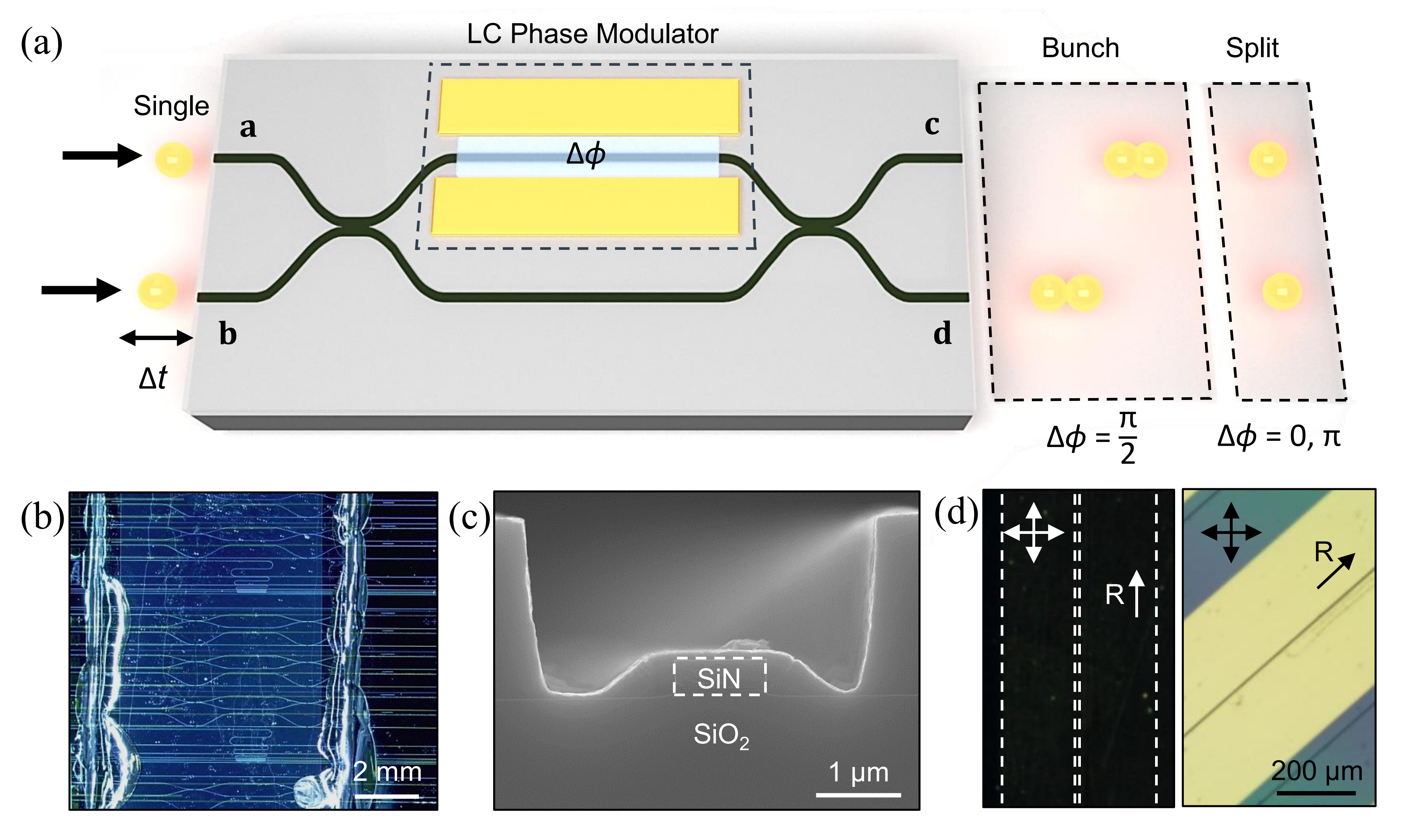}
\caption{(a) Schematic of two-photon NOON state generation in a MZI with an LC phase modulator. 
(b) Optical microscope image of the fabricated LC-MZI device. 
(c) SEM image showing a cross sectional view of the LC trench. 
(d) Polarized optical microscope images showing the LC alignment at the waveguide and electrode areas. 
The rubbing direction (R) is either parallel (left) or $45^{\circ}$ with respect to the polarizer (right). 
In the left image, white dashed lines illustrate the location of electrodes.}
\label{fig1}
\end{figure}

We demonstrate on-chip two-photon interference using a MZI composed of two beam splitters and a LC-based SiN phase modulator, as schematically depicted in \textbf{Figure \ref{fig1}(a)}. In our experiment, two photons are injected into the input ports (a and b), represented by the creation operators $a^\dagger$ and $b^\dagger$. The LC-MZI implements a unitary operation $U_{\text{LC-MZI}}$ that transforms the input state into a two-photon output state expressed in terms of the creation operators of the output modes $c^\dagger$ and $d^\dagger$. This output state is critically dependent on the phase difference ($\Delta\phi$) introduced by the on-chip LC modulator, as described by the following equation:

\begin{equation}
a^\dagger b^\dagger \xrightarrow{U_{LC-MZI}} \frac{1}{4} \left[ -i(e^{-2i\Delta\phi} - 1) c^{\dagger 2} - 2(e^{-2i\Delta\phi} + 1) c^\dagger d^\dagger + i(e^{-2i\Delta\phi} - 1) d^{\dagger 2} \right]
\end{equation}

This relation highlights the principle of our experiment: by electrically tuning the phase $\Delta\phi$, the quantum interference of the two photons at the output ports can be continuously controlled. Specifically, when the relative phase is set to $\Delta\phi = \pi/2$, the photons "bunch" and exit from the same output port (resulting in $c^{\dagger 2}$ or $d^{\dagger 2}$ events). In contrast, when $\Delta\phi = 0$ or $\pi$, they "split" and exit from different ports ($c^\dagger d^\dagger$). This voltage-controlled manipulation is achieved by the electro-optic response of the LC molecules, where the applied electric field induces molecular reorientation, modifying the effective refractive index of the waveguide mode. This represents the key operating principle of our integrated quantum platform, as schematically illustrated in \textbf{Figure \ref{fig1}(a)}.

For experimental implementation, we fabricated the device on a SiN on insulator platform. Silicon nitride was chosen for its wide transparency window and low propagation loss, which are crucial for preserving quantum states. \textbf{Figure \ref{fig1}(b)} shows an optical microscope image of the fully assembled LC-MZI device after LC injection. In contrast, the cross-sectional view captured by a scanning electron microscope (SEM) in \textbf{Figure \ref{fig1}(c)} reveals the device geometry prior to LC injection. Specifically, we designed the integrated photonic device on an LPCVD SiN layer with a height of $300~\text{nm}$, a full-etch depth of $300~\text{nm}$, and a waveguide width of $1~\mu\text{m}$, covered by a $2\text{-}\mu\text{m}$-thick PECVD SiO$_2$ cladding. To form the phase modulator, a trench with a width of $3~\mu\text{m}$ and a length of $500~\mu\text{m}$ was opened to maximize the overlap between the optical mode and the LC material, thereby enhancing the phase modulation efficiency. The successful injection and uniform alignment of the LC molecules, which is essential for precise and predictable phase modulation, was verified by crossed polarized optical microscopy, as shown in \textbf{Figure \ref{fig1}(d)}. This uniform alignment, driven by the anchoring force of the polyimide top alignment layer, ensures a consistent effective refractive index ($n_{\text{eff}}$), which is essential for implementing the high-fidelity, voltage-tunable phase shifter required for our two-photon interference experiments.

\subsection{Mechanism of the Liquid Crystal Phase Modulator}

\begin{figure}
\nolinenumbers
\centering
\includegraphics[width=0.9\columnwidth]{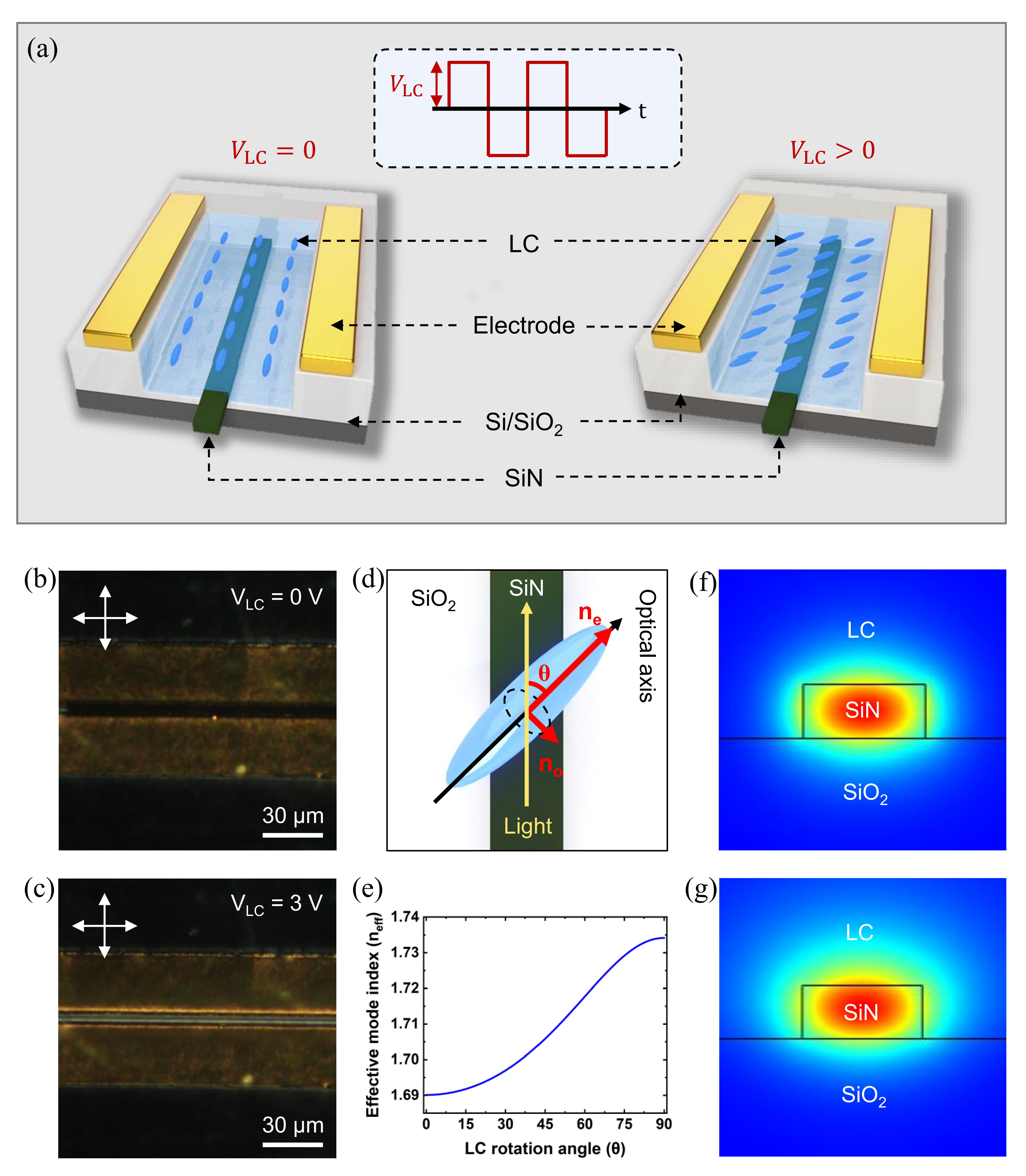}
\caption{(a) Enlarged view of the phase modulator illustrating the alignment of liquid crystal molecules driven by an applied square wave voltage. The left side shows the molecular alignment before applying voltage ($V_{\text{LC}} = 0$), while the right side shows the alignment after applying voltage ($V_{\text{LC}} > 0$). (b) Polarized optical microscope image of the phase modulator with liquid crystal molecules aligned before voltage application ($V_{\text{LC}} = 0$), corresponding to the left side of (a). (c) Polarized optical microscope image of the phase modulator with liquid crystal molecules aligned after voltage application ($V_{\text{LC}} = 3$ V), corresponding to the right side of (a). (d) Top-view illustration of liquid crystal molecules rotated by an angle $\theta$, with the waveguide passing beneath. The refractive indices of the liquid crystal are $n_e = 1.685$ and $n_o = 1.500$. (e) Graph showing the variation in $n_{\text{eff}}$ as a function of $\theta$ from (d). (f) Mode profile and $n_{\text{eff}} = 1.69$ at $\theta = 0^{\circ}$. (g) Mode profile and $n_{\text{eff}} = 1.73$ at $\theta = 90^{\circ}$.}
\label{fig2}
\end{figure}

\textbf{Figure \ref{fig2}(a)} illustrates the operational principle of the LC phase modulation section, emphasizing the alignment of LC molecules under the influence of the applied electric field. We used E7 (Merck) as the nematic liquid crystal material, which exhibits a stable nematic phase over a broad temperature range from approximately $-10 \,^{\circ}\mathrm{C}$ to $+58 \,^{\circ}\mathrm{C}$~\cite{ma2017refractive,deshmukh2008effect}. The left side shows the molecular alignment before applying voltage ($V_{\text{LC}}=0$), while the right side shows the reorientation after applying voltage ($V_{\text{LC}}>0$). To effectively modulate the LC molecules, an AC square wave is applied. DC voltage is avoided because it causes an ionic current, leading to ion accumulation at the electrodes~\cite{yang2014fundamentals}. This creates a screening effect that reduces the electric field acting on the LC molecules. We set the frequency to be approximately 500 Hz to 1 kHz to optimize the balance between the LC response and the prevention of ionic current.


We verified this voltage-dependent alignment change using a polarized optical microscope. This setup operates on the principle of cross-polarization, where the optical transmission is dictated by the birefringence-induced change in the polarization state. In the absence of an applied voltage ($V_{\text{LC}}=0$), the LC molecules align parallel to the polarizer. Consequently, the polarization state of the light remains unchanged, and the reflected light is completely blocked by the orthogonal analyzer, making the waveguide region between the two electrodes appear dark, as shown in \textbf{Figure \ref{fig2}(b)}. Upon application of 3 V, the electric field induces a reorientation of the LC molecules, creating a birefringence that alters the polarization of the light. This allows the reflected light to pass through the analyzer, making the LC area near the electrodes appear bright, as seen in \textbf{Figure \ref{fig2}(c)}. This visual evidence directly confirms the field-induced molecular reorientation that is fundamental to our modulation mechanism.

To investigate the optical response of the proposed modulator under an applied electric field, we performed finite element method simulations to model the molecular reorientation of the LC and its impact on the $n_{\text{eff}}$ of the fundamental TE guided mode. E7 has ordinary and extraordinary refractive indices of $n_o = 1.500$ and $n_e = 1.685$ at 1550 nm, respectively~\cite{tkachenko2006nematic}. As shown in \textbf{Figure \ref{fig2}(d)}, the LC orientation angle $\theta$ varies with the applied field. The effective index of the fundamental TE mode was then calculated as a function of this rotation angle, as plotted in \textbf{Figure \ref{fig2}(e)}. The simulations predict a total $n_{\text{eff}}$ change from 1.69 at $\theta = 0^{\circ}$ to 1.73 at $\theta = 90^{\circ}$, corresponding to a modulation depth of 0.04. These changes are visually corroborated by the simulated mode profiles shown in \textbf{Figure \ref{fig2}(f)} and \textbf{Figure \ref{fig2}(g)}, which illustrate the fundamental TE mode distributions corresponding to the initial ($n_{\text{eff}}=1.69$) and fully rotated ($n_{\text{eff}}=1.73$) states, respectively. For quantitative comparison, the thermo-optic coefficient of SiN is relatively modest ($2.45 \times 10^{-5}$~K$^{-1}$)~\cite{yong2022power}, making it difficult to achieve a comparable $\Delta n_{\text{eff}}$ through thermal modulation alone. This indicates that the LC-induced refractive index change represents a substantially larger and more efficient tuning range than that typically attainable with thermo-optic approaches.

\subsection{Performance Characterization of the Mach-Zehnder Interferometer}

\begin{figure}
\nolinenumbers
\centering
\includegraphics[width=1\columnwidth]{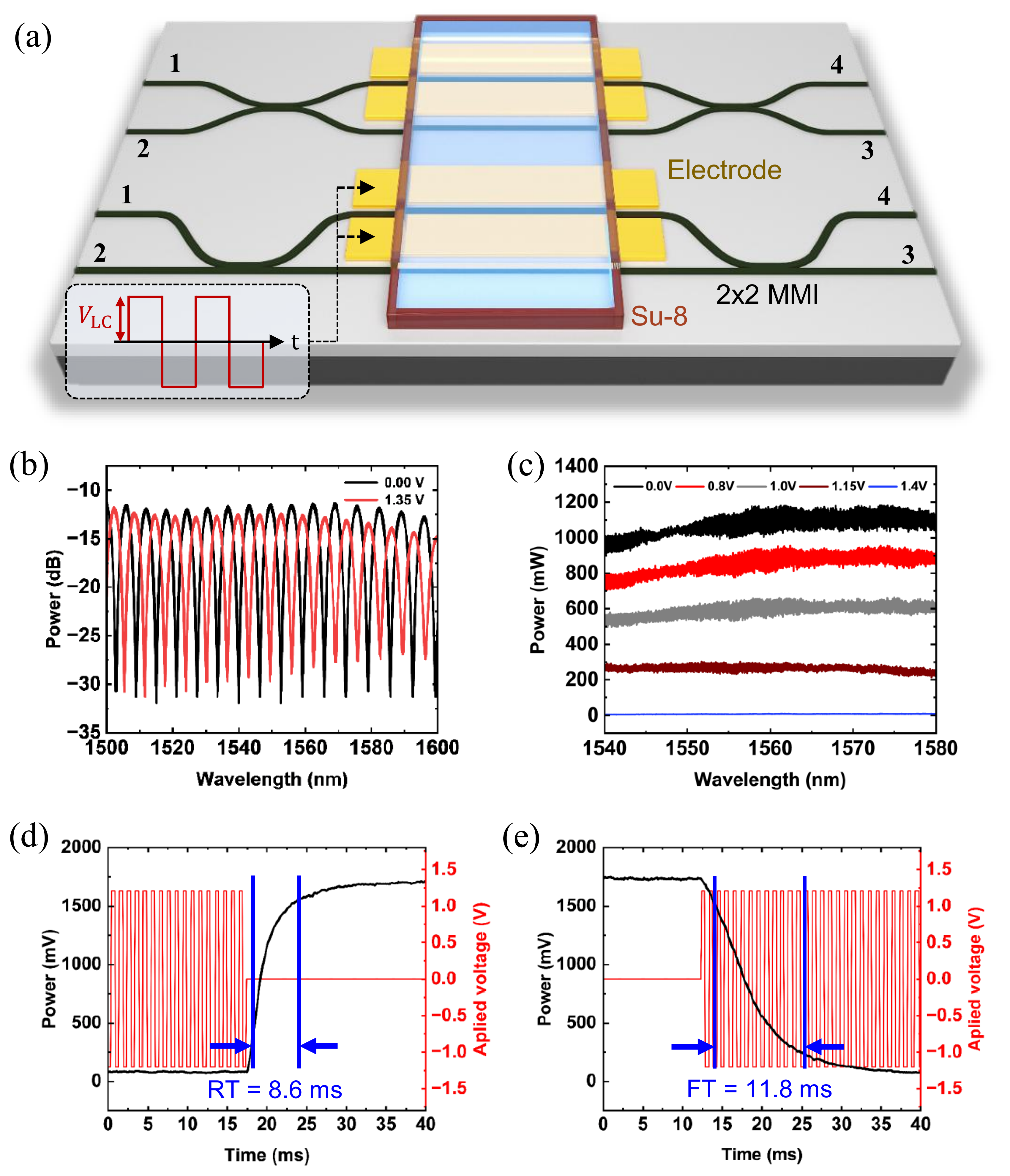}
\caption{(a) Illustration of the LC-MZI confined within SU-8 walls, consisting of two 2$\times$2 MMIs. The top figure shows a balanced and the bottom shows an unbalanced MZI. (b) Transmission spectra of the unbalanced MZI obtained by wavelength sweep, showing interference fringes at $V=0$ and at a voltage of $V_{\pi}=1.35~\mathrm{V_{LC}}$. (c) Transmission spectra of the balanced MZI, measured by wavelength sweep under increasing applied voltage from maximum to minimum transmission. The S32 parameter was measured to obtain these spectra. (d) Rise time (RT) and (e) fall time (FT) for an applied $V_{\pi}$ at $1~\mathrm{kHz}$, measured from the unbalanced MZI shown in (b).}
\label{fig3}
\end{figure}

To evaluate the performance of the LC-MZI modulator, we characterized both its static and dynamic properties. The static optical performance, encompassing the $V_{\pi}$, spectral tuning, and extinction ratio, was assessed using both balanced and unbalanced MZI configurations. Conversely, the dynamic temporal response was measured exclusively using the unbalanced MZI. As shown in \textbf{Figure \ref{fig3}(a)}, we fabricated both configurations, with the port numbers explicitly indicated to clarify the measurement setup.

The static performance was first evaluated by measuring the $V_{\pi}$. For the S$_{42}$ port of the unbalanced MZI, which incorporates a phase modulator length of $L = 500~\mu\mathrm{m}$, a full phase shift was achieved with a drive voltage of only $V_{\pi} = 1.35~\mathrm{V}$, as indicated by the fringe shift in \textbf{Figure \ref{fig3}(b)}. This low operating voltage confirms the device's compatibility with CMOS-level electronics and is consistent with our target of $V_{\pi} \cdot L < 1~\mathrm{V\cdot mm}$, validating the potential for energy-efficient operation in dense photonic arrays. 

To further visualize the device's response to an applied voltage, we measured the transmission spectrum of the balanced MZI's S$_{32}$ port. \textbf{Figure \ref{fig3}(c)} shows five overlaid spectra, measured at five voltages ranging from $0~\mathrm{V}$ up to the switching voltage of $V_{\pi} \approx 1.4~\mathrm{V}$. The results clearly demonstrate that the voltage-induced increase in the mode $n_{\text{eff}}$ causes the interference pattern to shift toward the longer wavelength (Red) Side. The measured $V_{\pi}$ of $1.4~\mathrm{V}$ is highly consistent with the $1.35~\mathrm{V}$ obtained from the unbalanced MZI, as this device also features the same modulator length of $L=500~\mu\mathrm{m}$. This consistency confirms the uniformity of the phase modulation performance across different device topologies. The spectral shift, combined with an observed extinction ratio of $\sim$20 dB between the maximum and minimum transmissions in the balanced MZI, confirms the device's voltage-controlled phase tuning capability and high fabrication quality.

Following the static characterization, the response times were measured from the S$_{32}$ port of the unbalanced MZI to assess the temporal limits of the modulation. A 1~kHz square wave voltage with an amplitude of $V_{\pi}$ was applied to determine the switching speed. Specifically, the measurement was conducted at a fixed wavelength of $1549.2~\mathrm{nm}$, corresponding to a constructive interference peak before applying voltage in \textbf{Figure \ref{fig3}(b)}. As plotted in \textbf{Figure \ref{fig3}(d)} and \textbf{Figure \ref{fig3}(e)}, the rise time (RT) and fall times (FT) were measured to be 8.6~ms and 11.8~ms, respectively. While these values are characteristic of nematic liquid crystal reorientation dynamics, they are sufficient for applications requiring circuit reconfiguration or active phase stabilization. Collectively, these static and dynamic characterizations demonstrate that our LC-MZI device meets the necessary performance benchmarks for implementing high-fidelity quantum interference experiments.

\subsection{Two-Photon Interference with Voltage-Controlled Phase Tuning}

\begin{figure}
\nolinenumbers
\centering
\includegraphics[width=\linewidth]{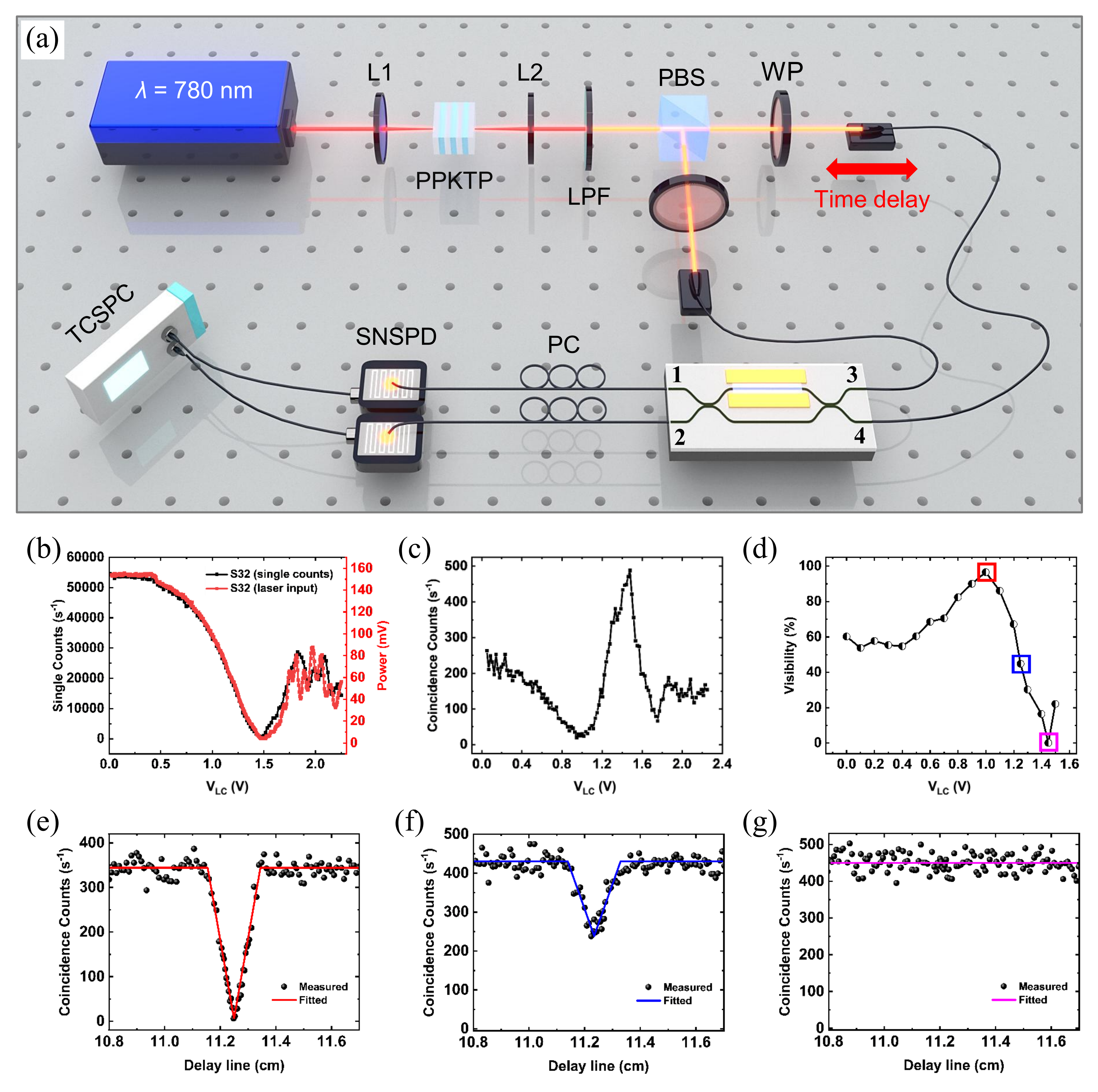}
\caption{(a) Experimental setup for two-photon interference. Photon pairs are generated in the optical system and injected into an integrated LC-MZI for phase-controlled interference.(b) Transmission at the cross port with a single-photon input injected into one arm of the MZI, compared with the classical transmission measured using a laser, showing similar dependence. (c) Coincidence counts measured with single-photon inputs at both arms as a function of applied voltage. (d) Visibility of two-photon interference under the same condition as (c). (e--g) Coincidence--time delay curves for different applied voltages: (e) visibility $\approx 98.5\%$ at $V = 1.06~\mathrm{V}$, (f) visibility $\approx 44\%$ at $V = 1.22~\mathrm{V}$, and (g) visibility $\approx 0\%$ at $V = 1.40~\mathrm{V}$. The solid lines in (e--g) represent fits to the experimental data.}
\label{fig4}
\end{figure}

To demonstrate the capability of our integrated LC modulator for quantum applications, we performed two-photon interference experiments. \textbf{Figure \ref{fig4}(a)} shows the experimental setup for generating photon pairs and performing two-photon interference on our integrated photonic chip. The foundation of this experiment is a high-quality source of photon pairs, which we produced via spontaneous parametric down-conversion in a 7 mm-long type-II periodically poled potassium titanyl phosphate (PPKTP) crystal pumped by a 780 nm continuous-wave (CW) tunable diode laser (Toptica, TA Pro). After separation by a polarizing beam splitter (PBS), residual pump light was rejected using long-pass filters. The polarization in each arm was independently prepared using a half-wave plate (HWP) and a quarter-wave plate (QWP). A translation stage in one path provided fine control of the relative delay, which is critical for observing Hong–Ou–Mandel (HOM) interference and verifying photon indistinguishability. The photons were coupled into and out of the chip using four lensed fibers and detected with superconducting nanowire single-photon detectors (SNSPDs). Polarization controllers (PCs) optimized the collection, and photon arrival times were recorded on a time-to-digital converter (quTAG, qutools) for time-correlated single-photon counting (TCSPC)-based coincidence analysis.

Before characterizing the device, we first verified the quality of the photon-pair source using a fiber beam splitter. The HOM effect, a fundamental two-photon interference phenomenon, was observed as shown in Figure S2(Supporting Information). The measured visibility, defined as 
\[
V = \frac{C_{\max} - C_{\min}}{C_{\max} + C_{\min}},
\]
after correcting for accidental coincidences, was 97.5\%. This high visibility confirms the high indistinguishability of the photon pairs and the reliability of the experimental setup.

Having verified the quality of the source with the fiber beam splitter, we proceeded to examine the on-chip response of the LC-MZI. First, we characterized the fundamental phase modulation response using a single input. \textbf{Figure \ref{fig4}(b)} displays the single-photon transmission at the cross port (S$_{32}$) as a function of the applied voltage. The transmitted optical power from a continuous-wave laser and the single-photon counts from a heralded single-photon source exhibit nearly identical modulation profiles, confirming that the LC phase modulator provides a consistent voltage-dependent phase shift regardless of the photon number statistics.

Next, we investigated voltage-controlled two-photon interference by injecting photon pairs into both input ports. To maximize the two-photon interaction, we fixed the delay line at the zero-path-difference position ($\tau=0$), ensuring that the photon pairs arrive at the MZI simultaneously. \textbf{Figure \ref{fig4}(c)} shows the two-photon coincidence counts recorded at this zero-delay point ($\tau=0$) while sweeping the applied voltage. The coincidence signal oscillates with twice the frequency of the single-photon transmission shown in \textbf{Figure \ref{fig4}(b)}, a hallmark of two-photon interference controlled by the LC modulator.

To further quantify this effect, we measured the quantum interference visibility by performing a full delay line sweep at each applied bias voltage. The resulting visibilities are plotted in \textbf{Figure \ref{fig4}(d)} as a function of the applied voltage, showing a maximum visibility of approximately 98.5\% at $V_{\mathrm{LC}} = 1.06~\mathrm{V}$. The corresponding HOM dip profiles measured at three representative voltages, 1.06~V, 1.22~V, and 1.40~V, are shown in \textbf{Figure \ref{fig4}(e)}, \textbf{Figure \ref{fig4}(f)}, and \textbf{Figure \ref{fig4}(g)}, respectively. These figures display the measured coincidence counts obtained by sweeping the delay line while maintaining $V_{\mathrm{LC}}$ at constant values, illustrating the variation in dip depth.

\subsection{Scalability and Large-Scale Production via Stepper Lithography}

While our previous devices fabricated using E-beam lithography successfully validated key quantum functionalities, such as two-photon interference, the process was plagued by critical issues with fabrication yield. This was primarily attributed to electrical instability and material degradation resulting from the direct contact between the electrodes and the LC. To resolve this yield problem and demonstrate scalability, we developed a new, robust fabrication process, now implemented on a 4-inch wafer scale using stepper lithography.

\begin{figure}
\nolinenumbers
\centering
\includegraphics[width=\linewidth]{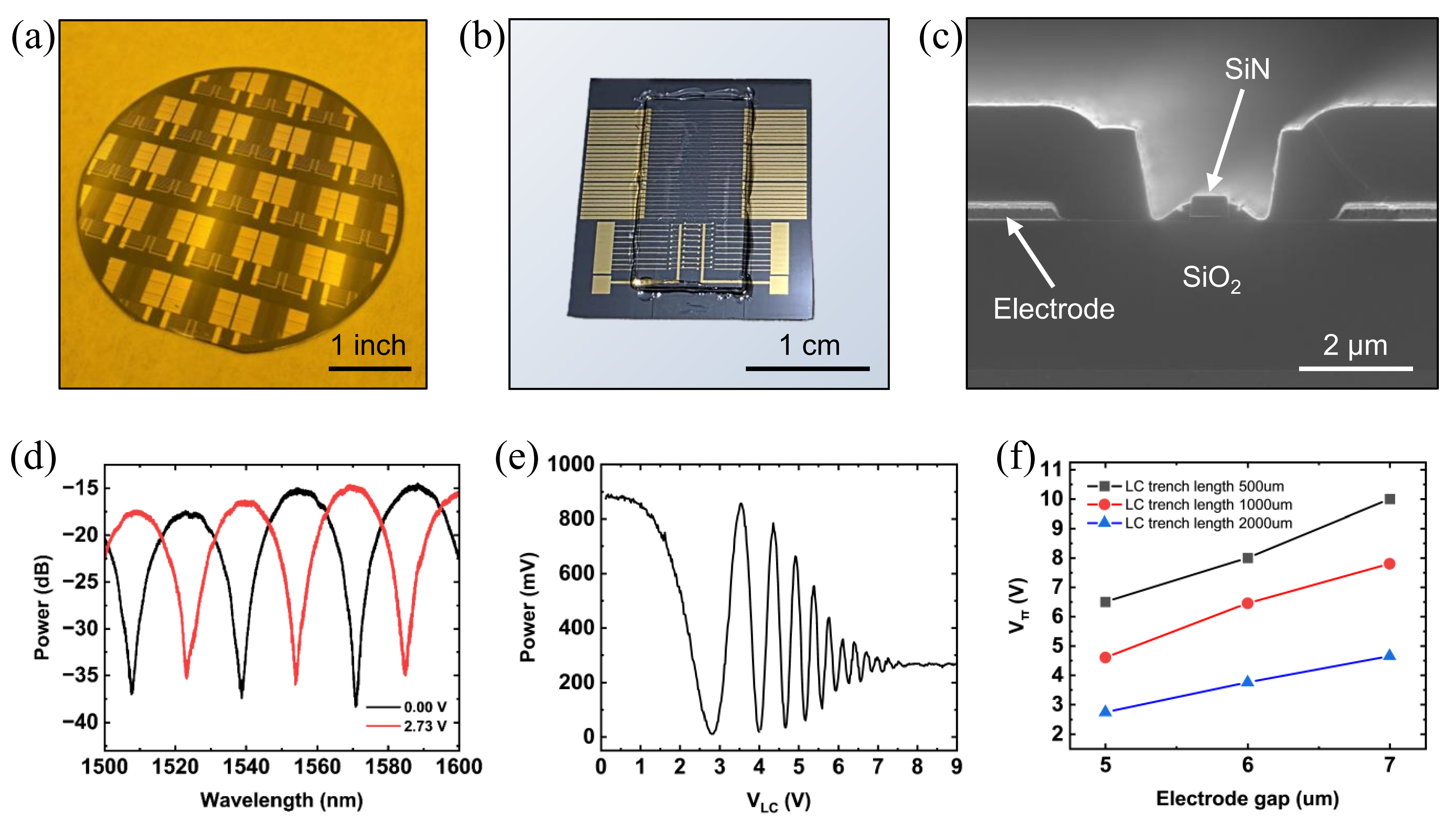}
\caption{(a) Wafer-scale image of the 4-inch wafer after completion of all fabrication steps (before LC injection). (b) Device after dicing, quartz bonding, and liquid crystal injection. (c) Scanning Electron Microscope (SEM) image of the liquid crystal trench region, showing the electrode gap and waveguide structure. (d) Transmission spectra of the unbalanced MZI with waveguides fabricated using stepper lithography, measured by wavelength sweep. (e) Transmission change versus voltage measured at a 1557 nm wavelength. (f) $V_{\pi}$ versus electrode gap for different LC modulator lengths.}
\label{fig5}
\end{figure}

The cornerstone of this new process is a design that structurally isolates the electrodes from the LC material. \textbf{Figure \ref{fig5}} summarizes the results. The successful fabrication on a 4-inch wafer is shown in \textbf{Figure \ref{fig5}(a)}, highlighting the uniformity and high yield achieved. A completed single device after dicing, bonding, and LC injection is shown in \textbf{Figure \ref{fig5}(b)}. Crucially, \textbf{Figure \ref{fig5}(c)} provides a detailed SEM view of this new architecture, clearly showing that the LC trench region is physically isolated from the electrodes. This isolation is the key to preventing electrical shorts and material degradation, thereby achieving the robust yield necessary for large-scale production.

The electro-optic characterization confirms the high performance of this reliable design. \textbf{Figure \ref{fig5}(d)} shows the transmission spectra of an unbalanced MZI for a device with a 2000~$\mu$m LC modulator length and a 5~$\mu$m electrode gap. The black line represents the spectrum at 0~V (before voltage application), while the red line shows the spectrum after applying $V_{\pi}$, confirming modulation via spectral shift. \textbf{Figure \ref{fig5}(e)} plots the normalized transmission versus voltage at 1557~nm. Finally, \textbf{Figure \ref{fig5}(f)} maps the $V_{\pi}$ as a function of the electrode gap (5, 6 and 7~$\mu$m) for different lengths of the LC modulator (500, 1000 and 2000~$\mu$m). The data clearly show that $V_{\pi}$ decreases as the electrode gap narrows and as the modulator length increases. This result provides key design parameters for optimizing low-power operation on this scalable and reliable platform.

\section{Conclusion and Discussion}

We have designed, fabricated, and demonstrated a low-voltage LC phase modulator integrated with a MZI for quantum photonic applications. This work represents the first experimental demonstration of voltage-controlled two-photon quantum interference in LC-integrated photonic circuits. The device exhibits a low $V_{\pi} \cdot L$ of $< 1~\mathrm{V\cdot mm}$, confirming CMOS-compatible operation, and preserves photon quantum coherence with an interference visibility of approximately 98.5\%. Furthermore, successful wafer-scale fabrication using stepper lithography validates the scalability and reconfigurability of this technology.

Building on this fundamental LC-MZI block, our approach provides a versatile foundation for scaling toward more complex quantum photonic architectures. The high performance demonstrated in a single modulator suggests that LC phase shifter arrays of compact, low-voltage LC phase shifters could enable multichannel reconfigurable circuits with dense integration~\cite{bogaerts2020programmable}. Such scalability would further highlight the advantages of our small-footprint modulators in realizing large-scale interferometric networks~\cite{bogaerts2020programmable}. In addition, low driving voltage opens the possibility of direct integration with custom ASIC-based electronic drivers, paving the way for co-integrated photonic and electronic systems~\cite{atabaki2018integrating,cai2021high}. Extending this approach to multichannel configurations will enable stable and programmable phase control for complex quantum state generation and manipulation~\cite{wang2018multidimensional}. These programmable components, even with relatively slow response times, can play a significant role in optical filtering, photonic quantum simulations, programmable Gaussian boson sampling, generation of resource states such as multi-photon cluster states or Gottestam-Kitaev-Preskill states, and large-scale interferometric architectures for fault-tolerant quantum computation~\cite{tzitrin2021fault,aghaee2025scaling,psiquantum2025manufacturable,larsen2025integrated,bao2023very,wang2018multidimensional}.

While this work employed the widely used nematic LC E7 on SiN to ensure scalability and fabrication compatibility, the same concept can be extended to other materials and device configurations. Incorporating ferroelectric or polymer-stabilized LCs could enable sub-microsecond or even gigahertz-range modulation~\cite{chiang2025ferroelectric,taghavi2025ghz}. Recent studies have also investigated ferroelectric nematic LCs as promising candidates for tunable and nonreciprocal quantum light generation in free-space configurations~\cite{sultanov2024tunable,pan2025nonreciprocal}. Integrating such ferroelectric nematic LCs with waveguide platforms could enable longer light–matter interaction lengths and facilitate their incorporation into complex interferometric architectures, thereby extending their applicability to on-chip quantum photonic systems. In parallel, heterogeneous integration via ink-jet or localized deposition techniques may allow spatially tailored optical functionalities to be realized within a single chip~\cite{alaman2016inkjet}. Moreover, combining LC tuning with intrinsically electro-optic materials such as lithium niobate or lithium tantalate could further enhance modulation bandwidth, paving the way for hybrid architectures that bridge large-scale low-speed programmability with high-speed operations~\cite{sun2020hybrid,weigel2016lightwave}. 

By demonstrating high-performance quantum photonic functionality on an LC-integrated platform, this work establishes a foundation upon which these advancements can be realized. Together, these developments position LC-integrated photonics as a scalable and energy-efficient platform for the next generation of quantum information technologies.

\section{Methods}

The entire device fabrication and LC cell assembly process is schematically illustrated in Figure S1(Supporting Information). The process began on a SiO$_2$ on Si wafer with a SiN layer. Waveguide patterns for the LC-MZI were defined using two lithography methods depending on the goal: electron beam lithography (JEOL, JBX-8100FS) for the quantum interference experiments (Figures 1-4) and i-line stepper lithography (Nikon, NSR-2005i10C) for the scalability demonstration (Figure 5). After patterning, the SiN waveguides were formed using a reactive ion etching (RIE) process, followed by the deposition of a silicon dioxide (SiO$_2$) upper cladding via Plasma-Enhanced Chemical Vapor Deposition (PECVD).

To enable interaction with the LC, several structures were created on the chip. First, a trench was opened in the SiO$_2$ cladding directly above the waveguide using a 405~nm maskless lithography system and a subsequent RIE process. Next, chromium/gold (Cr/Au) electrodes were deposited by e-beam evaporation and defined through a lift-off process. Finally, SU-8 photoresist was patterned using the maskless lithography system to form the microfluidic walls for the LC trench.

The final step was the LC cell assembly. A top quartz substrate was diced to match the dimensions of the SU-8 walls. A polyimide alignment layer was spin-coated onto the quartz and subsequently rubbed to define the molecular orientation for the LC. This prepared quartz lid was then aligned and bonded to the SU-8 structure using a UV-curable epoxy. Finally, the nematic liquid crystal E7 (Merck) was injected into the sealed cell cavity via capillary action to complete the device.

\medskip
\textbf{Acknowledgements}\par 
This work was supported by the National Research Foundation of Korea (NRF) (RS-2023-NR119925, RS-2024-00343768, RS-2025-25465174), the National Research Council of Science and Technology (NST) (GTL25011-000), and the Institute for Information \& Communications Technology Planning \& Evaluation (IITP) grant funded by the Korea government (MSIT) (RS-2025-02218723). This work was also supported by the Korea Institute of Science and Technology (KIST) research program (26E0001, 26E0011). The authors acknowledge the support provided by the National Information Society Agency (NIA) funded by the Ministry of Science and ICT (MSIT, Korea) [Quantum technology test verification and consulting support in 2024-2026].

\section*{Data availability}
The data that support the findings of this study are available from the corresponding authors upon reasonable request.

\section*{Author contributions}
C.A., S.-U.K., and H.K. conceived the experiments. 
C.A. and S.M. fabricated the devices. 
Y.H., S.L., and H.J.K. performed the liquid crystal injection and characterization. 
C.A. and J.L. established the experimental setup and performed the quantum measurements. 
H.J. provided essential resources and guided the experimental design. 
H.-Y.Y., S.-U.K., and H.K. supervised the project as corresponding authors. 
C.A. wrote the manuscript with input from all authors. 
All authors discussed the results and contributed to the final version of the manuscript.

\section*{Competing interests}
The authors declare no competing interests.
\clearpage
\section{References}
\bibliographystyle{naturemag_noURL}
\bibliography{reference}
\clearpage

\clearpage

\clearpage

\clearpage

\clearpage

\end{document}